\documentclass[12pt,preprint]{aastex}

\newcommand {\ea} {{\it et~al.}}
\newcommand {\be} {\begin{equation}}
\newcommand {\ee} {\end{equation}}

\shorttitle{Poynting flux in quasar jets}

\shortauthors{Sikora \ea}

\begin{document}

\title{Are quasar jets dominated by Poynting flux?}

\author{Marek~Sikora\altaffilmark{1},
Mitchell C.~Begelman\altaffilmark{2},
Greg~M. Madejski\altaffilmark{3,5}, and
Jean-Pierre Lasota\altaffilmark{4}
}

\altaffiltext{1}{Nicolaus Copernicus Astronomical Center,
Bartycka 18, 00-716 Warsaw, Poland; \tt{sikora@camk.edu.pl}}
\altaffiltext{2}{JILA, University of Colorado, Boulder, CO 80309-0440, USA}
\altaffiltext{3}{Stanford Linear Accelerator Center, 2575 Sand Hill Road,
Menlo Park, CA 94025, USA}
\altaffiltext{4}{Institut d'Astrophysique de Paris, 98bis boulevard Arago,
75014 Paris, France}
\altaffiltext{5}{Kavli Institute for Particle Astrophysics and Cosmology,
Stanford University, Stanford, CA 94305}

\begin{abstract}

The formation of relativistic astrophysical jets is presumably
mediated by magnetic fields threading accretion disks and central,
rapidly rotating objects. As it is accelerated by magnetic stresses,
the jet's kinetic energy flux grows at the expense of its Poynting
flux. However, it is unclear how efficient is the conversion from
magnetic to kinetic energy and whether there are any observational
signatures of this process.  We address this issue in the context
of jets in quasars. Using data from all spatial scales, we demonstrate
that in these objects the conversion from Poynting-flux-dominated
to matter-dominated jets is very likely to 
take place closer to the black hole than
the region where most of the Doppler boosted radiation observed 
in blazars is produced.  We
briefly discuss the possibility that blazar activity can be
induced by global MHD instabilities, e.g.,
via the production of localized velocity gradients that lead to
dissipative events such as shocks or magnetic reconnection, where
acceleration of relativistic particles and production of non-thermal
flares is taking place.

\end{abstract}


\section{INTRODUCTION}

The most promising scenario for launching astrophysical
relativistic jets involves large-scale magnetic fields anchored in
rapidly rotating compact objects.  The
idea of driving outflows by rotating magnetic fields, originally
invented by Weber and Davis (1967) to explain the spindown of
young stars, was successfully applied to pulsar winds (Michel
1969; Goldreich \& Julian 1970) and became a dominant mechanism in
theories of relativistic jets in active galactic nuclei (AGNs)
(see, e.g., Phinney 1983; Camenzind 1986, 1987; Li, Chiueh, \&
Begelman 1992; Vlahakis \& K\"onigl 2004) and gamma-ray bursts
(e.g., Thompson 1994; Spruit, Daigne, \& Drenkhahn 2001; Vlahakis
\& K\"onigl 2001). Powerful jets in quasars can be powered by 
innermost portions of an accretion disks and/or by rapidly
rotating black holes. They  can become relativistic if the total
to rest-mass energy flux ratio $\mu \equiv L_j / \dot M c^2 \gg
1$, where $L_j= L_B + L_K$ is the total energy flux, $L_B$ is the
magnetic energy flux, $L_K=(\Gamma-1)\dot M c^2$ is the kinetic
energy flux, and $\dot M$ is the mass loading rate.

Theories of axisymmetric, steady-state ideal-MHD jets predict that
the conversion process transforming electromagnetic energy
into plasma kinetic energy works efficiently only up to the
classical fast-magnetosonic surface, $z_{fm}$, located at a few
light cylinder-radii (Sakurai 1985; Li et al. 1992; Beskin,
Kuznetsova, \& Rafikov 1998). At this distance, the ratio of
Poynting flux to kinetic energy flux, $\sigma$, drops  to the
value $\sim \mu^{2/3}$.
This means that for $\mu \gg 1$ the flow still remains
strongly Poynting flux-dominated at $z_{fm}$.
Whether and how fast the conversion
can proceed beyond this point is unclear, firstly because of theoretical and
numerical difficulties in treating the strongly nonlinear differential
equation which combines the evolution of the flow along the magnetic surfaces
(Bernoulli equation) and the
cross-field-force balance (Grad-Shafranov equation), and secondly because
of the very uncertain boundary conditions at the base of the outflow and
along the outermost magnetic surface
(Begelman 1998; Heyvaerts \& Norman 2003; Vlahakis \& K\"onigl 2003a, b).

Signatures of the presence of dynamically strong, ordered  magnetic fields
have been pursued observationally, by tracing magnetic field structure
(using polarization information)
and kinematics of  kiloparsec- and parsec-scale jets and by
studying morphology of extended radio sources.
The results of these efforts are critically
reviewed in \S2. We demonstrate that
there is no convincing evidence for dynamically important
magnetic fields in quasar jets on scales larger than a few parsecs.
In \S 3, we extend the above studies down to subparsec regions, by
using  multi-wavelength observations of blazars. In particular,
the X- and $\gamma$-ray data allow us to estimate the leptonic and
total energy flux, which in turn provides an upper limit for the magnetic
dominance and a lower limit for the pair content.   Finally, in \S4,
we investigate constraints imposed on the jet dynamics by
the lack of signatures of the bulk Compton radiation in the blazar spectra.
Implications of our results for the global jet dynamics and for jet
launching scenarios are discussed in \S5.

\section{MAGNETIC FIELDS IN RADIO JETS}

\subsection{Kiloparsec-scale jets}
Arguments used in favor of the dynamical dominance of magnetic
fields over large spatial scales include the high linear
polarization of kiloparsec-scale jets, and the need for {\it in
situ} energy dissipation to provide  the power for acceleration of
fast-cooling ultra-relativistic electrons/positrons responsible
for synchrotron radiation in the optical band, and in some objects
also in the X-ray band  (see, e.g., Lesch \& Birk 1998; Birk \&
Lesch 2000). However, neither the energy dissipation nor the high
polarization require large scale mean magnetic fields: energy can
be dissipated in shocks and in  boundary shear layers and high
polarization can result from compression and stretching of
initially tangled/turbulent magnetic fields (Laing 1980, 1981;
Cawthorne et al. 1993). Furthermore, any large scale mean magnetic
fields on parsec and larger scales are expected to be strongly
dominated by the toroidal component, even if they started as
purely poloidal. This arises from the fact that due to
conservation of the magnetic flux, the poloidal component drops
with distance $\propto 1/r^2$, while due to conservation of the
magnetic {\it energy}  flux, the toroidal component drops $\propto
1/r$, where $r$ is the distance in a conical jet (Begelman,
Blandford, \& Rees 1984).  This is in contrast to inferences based
on polarimetry of large-scale jets in FRII radio galaxies and
quasars, which indicate a  parallel magnetic field orientation
(Bridle et al. 1994). Such magnetic fields can be generated by a dynamo
in the turbulent shear layers (Urpin 2002). The turbulence, driven by 
a shear, can be also responsible for acceleration of relativistic
electrons producing nonthermal radiation in kiloparsec scale jets 
(Stawarz \& Ostrowski 2002). Direct support for
such a scenario is provided by measurements  of intensity
and polarization profiles across jets in a number of nearby FRII
radiogalaxies (see Swain, Bridle, \& Baum 1998, and references
therein). There, the lack or deficit of radiation from the jet
spine can result from the Doppler deboosting, because the flow in
the jet spines can move faster than the flow in the
boundary/shearing layers. Indeed, these spines are presumably the
regions that produce X-ray radiation seen in some large-scale
quasar jets oriented at small angles to the line of sight. Models
of this radiation predict jet Lorentz factors significantly larger
than those derived from the radio brightness asymmetries of jets
and counter-jets (Tavecchio et al. 2004).

The hydrodynamical nature  of large  scale jets is
also indicated by numerical simulations of their termination. As
was demonstrated by Clarke,
Norman, \& Burns (1986) and Lind et al. (1989) in the  non-relativistic case,
and by Komissarov (1999) in the relativistic case, jets with  strong
toroidal magnetic components do not develop substantial back-flowing cocoons.
Instead, the shocked jet plasma, being confined by magnetic stresses, forms
a ``nose cone''--shaped head.  The cocoons observed in classical FR II radio
sources do not appear to have such nose-cones. They are broad and their
morphologies agree very well with those predicted theoretically
(Scheuer 1974; Begelman \& Cioffi 1989) and confirmed via numerical
simulations for light, supersonic, unmagnetized jets.
It should be noted, however, that all numerical simulations of strongly
magnetized jets have been performed assuming ideal MHD flows, whereas
the current closure condition requires a large fraction of 
magnetic fields to be resistively dissipated
(Benford 1978; Lesch, Appl, \& Camenzind 1989). Whether this can lead to
formation of very compact hot spots observed in radio lobes remains very
unclear, particularly because no  strong terminal shocks are expected 
to be produced by jets with
$\sigma >1$.

\subsection{Parsec-scale jets}

As was  pointed out by Blandford (1993), the toroidal component of
magnetic fields in Poynting flux-dominated jets can be indicated
by gradients of the Faraday rotation across parsec-scale jets.
Such gradients have been found in several BL Lac objects (Gabuzda,
Murray, \& Cronin 2004) and in quasar 3C273 (Asada et al. 2002).
However, the quality of these data, particularly regarding 3C273,
is still very poor due to limited resolution of the transverse jet
structure. And, as a strong counterexample, we refer to the only
quasar for which there is reasonably good transverse resolution,
B1611+343. No Faraday rotation gradient is seen in this object
(Zavala \& Taylor 2003). Furthermore, Faraday rotation can be
produced in an external non-uniform medium, provided that such a
medium is located in close proximity to the jet as required by 
$\sim 1$ year variations of the rotation measure in 3C 273 and 3C
279 (Zavala \& Taylor 2001). The external origin of Faraday
rotation is strongly indicated by the fact that in most objects
the rotation follows the $\lambda^2$-rule, where $\lambda$ is the
wavelength of the electromagnetic radiation (Gabuzda et al. 2004;
Zavala \& Taylor 2004). That rule is followed
even in objects with  rotation exceeding 1 radian, which cannot be the case 
if produced co-spatially with the radiation (Burn 1966; 
Gardner \& Whiteoak 1966).
Furthermore, observations of optically thick radio cores at different 
wavelengths indicate that position of the  polarization angle is the same
at different synchrotron photospheres, which indicates that it doesn't depend 
on a distance from the center (Zavala \& Taylor 2004). This can be reconciled
with theoretically predicted strong dependence of the rotation measure
on a distance in the laterally expanding jet ($RM \propto 1/r^2$ for conical 
jets), only if the Faraday depth of the parsec scale jets is negligible.

Arguments based on kinematics are also used in the literature to
argue in favor of the dynamical domination of the magnetic fields
in parsec-scale jets. The detection of any systematic acceleration
of the flow would provide a strong argument for the conversion of
the Poynting energy into kinetic energy.  Homan et al. (2001)
claim that such acceleration is indicated by VLBI observations of
sources having multiple components with proper motion. They point
out that in such objects the innermost components are
significantly slower than the others. However, within the same
sample of sources, the authors are not able to identify any clear
case of individual components that are speeding up. Moreover, the
above claims about slower-moving innermost components seem to
contradict the finding that there is a systematic decrease in
apparent velocity with increasing wavelength (Kellermann et al.
2004). The simplest interpretation of this is that the
observations at longer wavelengths cover more extended portions of
the jet structure, and therefore that the radio components
decelerate, rather than accelerate. Alternatively, it is possible
that a jet possesses a transverse structure and that the
central portions of the jet, which presumably move faster than the
boundary layers, contribute more radiation at the shorter
wavelengths.  Noting also that some jets appear curved, one
shouldn't be surprised to see sometimes both increasing and
decreasing projected speeds (see, e.g., Jorstad et al. 2004). It
should also be emphasized that if such sources are intrinsically
expanding, their surface-brightness peaks do not represent the
real component centers and the apparent offset can change with
time. In this case, one learns little about the intrinsic
kinematics of the source from the motion of the
surface-brightness-peak of the radio component. Furthermore, the
features that appear as moving on the VLBI scale may represent
moving patterns, rather then the real flow speeds. This is clearly
documented, for example,  by observations of a parsec-scale jet in
Mrk501.  There, the one-sidedness  of the jet and the lack of
superluminal motion of the VLBI features strongly suggest that the
the knots represent  stationary or very slowly moving patterns,
presumably oblique shocks (Giroletti et al. 2004; Ghisellini,
Tavecchio, \& Chiaberge 2004). Finally, even if some apparent
acceleration events are real, they are not necessarily related to
the conversion of magnetic energy to kinetic energy.  Acceleration
events can be produced also in matter-dominated jets, e.g., at the
expense of internal energy dissipated in shocks and partially
 returned to the flow, or due to the collision of the flow with an external
cloud.  Given the above, we regard as somewhat premature the
claims that two ``accelerating'' individual features, in 3C 279
(Piner et al. 2003) and 3C 345 (Unwin et al. 1997; Lobanov \&
Zensus 1999), respectively, prove the magnetic domination of
parsec-scale jets in these objects as asserted e.g. by
(Vlahakis \& K\"onigl 2004).

Another approach to studying the dynamics of a jet is based on comparing its
surface brightness distribution with that of its counterjet.  This method was
applied by Sudou et al. (2002) to the parsec-scale two-sided jet in the radio
galaxy NGC 6251. They found that the jet--to--counterjet intensity ratio
increases with distance from the center and using these data
derived the acceleration of the flow, with $v \simeq 0.13c$ at a distance
$r \simeq 0.53$pc and $v \simeq 0.42c$ at $r \simeq 1$pc. However, the reality
of the counterjet detection  in NGC 6251 is questioned by Jones \& Wehrle
(2002). In particular, they did not confirm the  presence of the conterjet at
$15$ GHz, the frequency at which their observations  had
a similar angular resolution to  those by Sudou et al. (2002).

Both methods, involving the jet/counterjet brightness ratio
and the apparent speed of the radio-emitting features, have been
used by Cotton et al. (1999) to derive the  dynamics of a parsec-scale jet
in radio galaxy NGC  315. They infer an acceleration  from $0.77c$ at
a distance 3.3 pc to $0.95c$ at a distance 9.5 pc.
However, as the authors point out, the present data are insufficient
to determine
whether the observed emission is from the main body of a jet or from
its slower outer layers.

\section{STRUCTURE OF ``BLAZAR JETS''}

\subsection{Polarization properties and dissipative events}

Short variability timescales --- of the order of $1$ week in the optical band
and similar or even shorter with larger amplitudes in the
$\gamma$-ray band (von Montigny et al. 1995; Mukherjee et al. 1997) ---
show that most of the non-thermal radiation from blazars is produced in
a region
with a  size $R \le 10^{17} (t_{fl}/3 \, {\rm days}) (\Gamma/15)$ cm,
too compact to be transparent in the radio band (Sikora et al. 1994).
Such a compactness, combined with the transparency of blazars
to high-energy $\gamma$-rays and the lack of $\gamma$-radiation from
the radio lobe-dominated quasars,
implies that most of the high-energy radiation from blazars originates
in well-collimated  and relativistic (sub)parsec-scale jets.

Polarimetry measurements  of the variable optical, infrared and mm radiation
suggest that at subparsec distances, magnetic fields are dominated by
the transverse component (Impey, Lawrence, \& Tapia 1991;
Gabuzda \& Sitko 1994; Cawthorne \& Gabuzda 1996;
Stevens, Robson, \& Holland 1996; Nartallo et al. 1998).  Such an orientation
is consistent with a toroidal magnetic field geometry, but can also result
from compression of a tangled magnetic field in transverse internal shocks.
The internal shocks have been proposed to result from collisions between
velocity inhomogeneities propagating down a matter-dominated jet
(Sikora et al. 1994; Spada et al. 2001).
The internal shock scenario
seems to be supported by the very broad energy distributions of relativistic
electrons/positrons. They cover 3-4 decades in energy and are injected with
approximately equal amounts of energy per decade (Moderski et al. 2003).
This contrasts strongly with the narrow energy distributions of accelerated
electrons predicted by the magnetic  reconnection models
(Zenitani \& Hoshino 2001; Larrabee, Lovelace, \& Romanova 2003).
However, it should be emphasized here that acceleration mechanisms
for electrons/positrons still await quantitative theories, with
regard to both reconnection and shock scenarios, if the inertia of
the matter is dominated by protons.

\subsection{Leptonic flux}

Shortly after the discovery of variable and strong $\gamma$-rays from blazars
by the {\it Compton Gamma-Ray Observatory} (CGRO), it was realized that
the exceptionally high $\gamma$-ray luminosities of quasar-hosted blazars
can result from
Comptonization of external radiation fields by ultra-relativistic
electrons/positrons injected in sub-parsec scale jets propagating with
a bulk Lorentz factor  $\Gamma \sim 10-20$
(Dermer, Schlickeiser \& Mastichiadis 1992;
Dermer \& Schlickeiser 1993; Sikora, Begelman, \& Rees 1994;
Blandford \& Levinson 1995).  Indeed, the data collected during the entire
period of the CGRO mission strongly support this idea.  All main features of
the high-energy spectra of blazars during $\gamma$-ray flares
can be explained naturally in terms of the external-radiation-Compton
(ERC) model (Moderski, Sikora, \& B{\l}a\.zejowski 2003). In particular,
the distances at which short-duration flares are produced,
inferred from their variability time-scales ($t_{fl}$) to be
$z_{fl} \simeq c t_{fl} \Gamma^2 \simeq
1.7 \times 10^{18} (t_{fl}/3 d) (\Gamma/15)^2$ cm,
agree with the distance estimates obtained assuming that high-energy
spectra of blazar flares are produced by Comptonization of  broad emission
lines and that the spectral break between the hard X-ray and soft $\gamma$-ray
spectral ranges, observed in the 1 -- 30 MeV range, is due to cooling effects.

Spectra below the break are hard, with a power-law index of
the energy flux, $F_{\nu}^{-\alpha}$, in the range $0.3 \leq
\alpha \leq 0.8$ (Reeves \& Turner 2000; Donato et al. 2001;
Giommi et al. 2002). In the mid/soft X-ray bands the spectra often
soften, presumably due to the contribution from the
synchrotron-self-Compton (SSC) component (Sikora et al. 1994; Kubo
et al. 1998). However, there are several blazars for which the
spectra are well-fit by a single, very hard power-law function over
all  X-ray bands (Tavecchio et al. 2000).
These objects provide an
exceptional opportunity to study the energy distribution of
relativistic electrons/positrons down to their lowest energies.
(Note that the radiation by such electrons cannot be detected
in the synchrotron and SSC spectral components, in the
former because of synchrotron-self-absorption, in the latter
because the low-energy tail of the SSC spectrum is hidden by the
synchrotron component.) If X-rays in these blazars are indeed
produced by Comptonization of broad emission line light, then the
X-ray spectra  provide evidence that there is no low-energy
cutoff in the energy distribution down to mildly relativistic
energies.

Since even the hardest X-ray spectra in blazars have radiation energy indices
$\alpha > 0$, the total number of electrons and positrons involved in
an outburst, $N_e$, is well determined by soft/mid X-ray observations. This is
because, for a number distribution $N_\gamma \propto \gamma^{-s}$, with
$s=2\alpha +1 >1$,  $N_e = \int_{\gamma_{min}} N_{\gamma} \, d\gamma$ is
insensitive to the upper limit of the integral and the details of the
 particle distribution at higher energies.
The most stringent limit on the particle content of the jet
arises from considerations of the bulk-Compton radiation,
where the ``cold'' electrons carried along in the jet Compton-upscatter
the broad emission line photons. For an apparent luminosity
$\nu_{sx} L_{\nu_{sx}}$, determined  at
$h\nu_{sx} = h\nu(\gamma_{min} \sim 1) \sim 2$ ${\rm keV}(\Gamma/15)^2
(h\nu_{diff}/10$ $eV)$, one finds that
the electron number flux is (Sikora \& Madejski 2000)
\be \dot N_{e} \sim {N_e \over t_{fl}} \sim {\nu_{sx}L_{\nu_{sx}} \over
c \sigma_T u_{BEL} t_{fl} {\Gamma}^6} \sim
 10^{50} {(\nu_{sx}L_{\nu_{sx}})_{46} \over
(t_{fl}/3d)(\Gamma/15)^6 } \  {\rm s}^{-1} \, , \label{ele1}
\ee
where $u_{BEL} \simeq 3 \times 10^{-3}$ erg cm$^{-3}$ s$^{-1}$ is
the radiation energy density in the broad emission line region (Kaspi et al.
2000) and $\nu_{sx}L_{\nu_{sx}} \equiv
10^{46} \times  (\nu_{sx}L_{\nu_{sx}})_{46}  \ {\rm erg \  s}^{-1} $.

The leptonic flux can be also estimated indirectly, using  data from
the $\gamma$-ray band where radiation is produced in the fast cooling regime.
Assuming a single power-law electron injection function, with
$\gamma_{min} \sim 1$ and
the index $p = 2$ corresponding to the high energy $\gamma$-ray spectrum
slope $\alpha_{\gamma} \simeq 1$ (Pohl et al. 1997),
we have (see Eq.12 in Moderski et al. 2004)
\be \dot N_e \sim  10^{51}
{(\nu L_{\nu})_{48} \over (\Gamma /15)^3 }\ {\rm s}^{-1}
, \, \label{ele2} \ee
where $(\nu L_{\nu})_{48} \equiv (\nu L_{\nu})/10^{48}$erg s$^{-1}$
is the apparent luminosity of the radiation produced at some frequency
in the fast cooling regime.

\subsection{Energetics}

Modeling of the observed high-energy spectra in blazars
within the context of the ERC model
gives the average
energy of the accelerated electrons/positrons $\bar \gamma_{inj} \sim 10$.
Heating the plasma by such a factor requires the total energy flux prior to
the dissipative event to be at least $\bar \gamma_{inj}$ times larger than
\be L_e = \dot N_e m_e c^2 \Gamma \simeq 10^{45} \dot N_{e,50} (\Gamma/15) \
{\rm erg \ s}^{-1} \, . \label{Le} \ee
In principle, this energy could be provided by the bulk
motion of electrons/positrons and dissipated via shocks formed due
to interactions with the external medium. This, however,
would require the initial bulk Lorentz factor to be $\Gamma_0 \sim
\bar \gamma_{inj}\Gamma \sim 150$.
Such a relativistic
e$^+$e$^-$-beam would produce a very strong spectral peak around
$\Gamma_0^2 \, h\nu_{UV} \sim 200$ keV due to Compton boosting  of
the external UV photons by a factor $\Gamma_0^2$. No such peaks
have been observed so far. In addition, electron/positron
jets, even if they can be successfully produced with $\Gamma_0 >
100$, would be immediately decelerated down to $\Gamma <10$ by
radiation drag in the compact central region
(Phinney 1987; Sikora et al. 1996). The
dissipative events cannot be powered by internal shocks, either,
if the flow consists purely of pair plasma. The internal shocks
can be, at most, mildly relativistic, even for very large ratios
of the bulk Lorentz factors of the colliding inhomogeneities
(Komissarov \& Falle 1997; Moderski et al. 2004). Therefore, the
acceleration of electrons/positrons to energies higher than $\bar
\gamma_{e,inj} \sim  2$ by shocks requires the energy flux in the flow to be
dominated by protons. Alternatively, if the energy flux in a jet
is dominated by magnetic fields, the pairs can be accelerated at
sites of magnetic field reconnection. In both cases, the jet
power can be estimated using the formula
\be L_j \simeq {\bar \gamma_{inj} L_e \over \eta_{diss} \eta_e}
\sim 10^{47} {(\bar \gamma_{inj}/10) L_{e,45} \over
(\eta_{diss}\eta_e/0.1)} \ {\rm erg \ s}^{-1}
\, , \label{Lj} \ee
where $\eta_{diss}$ is the efficiency of the energy dissipation in
the blazar zone (the region where most of the blazar radiation is
produced) and $\eta_e$ is the fraction of the dissipated energy
converted to relativistic electrons.
Jet powers inferred from the energetics of radio lobes
and hot spots cluster around $10^{46}$ erg s$^{-1}$ (Rawlings \&
Saunders 1991; Willott et al. 1999; D'Elia, Padovani \& Landt
2003; Maraschi \& Tavecchio 2003). These estimates are mutually
consistent, considering that the latter value is time-averaged
whereas the blazar energetics is determined from flares.

\subsection{Pair content}

Assuming that the energy flux of protons $L_p \simeq \dot N_p m_p c^2 \Gamma$
is $\gg L_e$ one can find that the pair content of quasar jets is
$n_{pairs}/n_p = (n_e/n_p - 1)/2$, where
\be {n_e \over n_p} \equiv {\dot N_e \over \dot N_p} =
{m_p \over m_e} {L_e \over L_j} (1 + \sigma)
\simeq 20 (1 + \sigma) {(\eta_{diss}\eta_e/0.1) \over (\bar \gamma_{inj}/10)}
\, ,\label{nenp} \ee
where $\sigma \simeq L_B/L_p$. The pair content is minimal for
$\sigma \ll 1$, with $n_e/n_p \sim 20$.  In this case $L_p \gg L_B$ and
the blazar dissipative events are likely to be dominated by internal shocks.

For $L_B > L_p$, shocks cannot be sufficiently
strong to provide efficient
particle acceleration and, therefore,
the dissipative events are more likely to originate via the
reconnection of magnetic field.
Since the minimum kinetic energy flux is given by $L_e$ {(cf. Eq. 3)},
jets which are most severely dominated by magnetic field would
have,  prior to the blazar dissipative zone,
$\sigma \simeq L_B/L_e \simeq 100 L_{j,47}/L_{e,45}$.
In these jets   $L_p < L_e$, i.e.,  $n_p < (m_e/m_p) n_e$.
Such jets would remain strongly Poynting-flux-dominated even after passing
through the dissipative blazar zone, with
$\sigma \simeq 10 L_{j,47}/[(\bar \gamma_{inj}/10) L_{e,45}]$.

\section{WITHIN THE FIRST $10^3-10^4$ GRAVITATIONAL RADII}

\subsection{Bulk-Compton radiation}

Another interesting constraint provided by blazar observations concerns
the efficiency of the jet acceleration process. Electrons/positrons,
dragged
through the very central region by protons and/or magnetic fields with bulk
Lorentz factor $\Gamma \ge 5$, would Comptonize the direct radiation from the
accretion disk, producing a strong soft X-ray bump (Begelman \& Sikora 1987;
Sikora \& Madejski 2000), or, in the case of
a non-stationary flow, soft X-ray precursors of the non-thermal flares
(Moderski et al. 2004).  The absence of such features in blazar spectra
indicates that the acceleration of jets up to $\Gamma \ge 10$ must take
at least $\sim 10^3$ gravitational radii.

\subsection{The conversion distance}

If blazar activity is related to the internal shocks formed in
unsteady, matter-dominated jets, the efficient energy dissipation
and particle acceleration imply $\sigma \ll 1$. This means that
the magnetic-to-kinetic energy conversion of the flow should take
place within a distance $z_{fl} \sim (t_{fl}/3 d) (\Gamma/15)^2$
pc. Is this achievable by an ideal MHD flow? In some
circumstances, acceleration by the magnetic nozzle effect can
work:  this can give $\Gamma/\Gamma_{fm} \sim \ln(z/z_{fm})$,
where $\Gamma_{fm} \sim \mu^{1/3}$ is the Lorentz factor at the
fast magnetosonic surface located at a distance $z_{fm} \sim
\mu^{2/3} z_c$, and $z_c$ is the distance at which a given
magnetic flux tube crosses the corresponding light-cylinder (Li et
al. 1992; Begelman \& Li 1994; Beskin, Kuznetsova, \& Rafikov
1998). From the assumption that in the blazar zone the jet is
dynamically dominated by protons, we have $\mu \simeq L_p/\dot M_j
c^2 \simeq \Gamma$ and find that the jet conversion can eventually
occur at a distance
\be z_{conv} \sim z_{fm} \exp(\Gamma/\Gamma_{fm}) \sim
(z_c/30R_g) M_9 \, {\rm pc} \, . \ee
This result ($z_{conv} \sim z_{fl}$) may indicate a physical
connection between the jet conversion process and dissipative
processes responsible for flaring activity of blazars. Such an activity
is likely to be induced by pinch and kink instabilities, developed
in the phase when a jet is still dominated by the Poynting flux
(Eichler 1993; Begelman 1998). MHD
instabilities can differentiate the flow speed and this in turn
can lead to the formation of shocks. An advantage of such a
scenario is that it avoids the need for modulation of the outflow
by the central engine. 
Any modulation of powerful jets at the base
would require strong instabilities in the main body of the
accretion disk (the corona is too tenuous to affect the jet
dynamically) which in turn are very likely to produce rapid
($\sim$ few days) and high amplitude variations of the UV-bump.
However, no such variations are indicated by optical observations
of nuclei of radio lobe-dominated  quasars (Stalin et al. 2004).
(We would like to comment here, that short term flaring activities are 
observed in blazars during both high states and low states 
(see, e.g., Webb et al. 1990), and that small  average amplitudes 
of short term variations can result from contamination of flares 
by radiation from larger distances in a jet or from superposition
of a varying number of flares observed at once.
An important factor determining  brightness of individual flares can be
related to the non-axisymmetry of the kink instability, which may cause  
the matter involved in different dissipative events to move in somewhat 
different directions relative to the line of sight.)
  
Alternatively, blazar activity can be directly related to magnetic energy
dissipation processes, without involving shock formation.
The largest uncertainty with this scenario is related
to two questions: whether it can lead to acceleration of electrons with
a very broad energy distribution, as indicated by the blazar spectra;
and whether in the acceleration sites the magnetic field geometry is dominated
by the transverse  component, as indicated by the  blazar polarimetry.
Since the magnetic energy dissipation can involve fragmentation of
long  current sheets into nonlinear  smaller structures of all sizes,
generation of broad power-law electron spectra may  be feasible, but
an important question still remains:
will enough of the transverse magnetic field orientation persist in order
to provide linear optical polarization, which in some objects can
reach a value up
to 20-40 \% and is preferentially aligned with a jet?

\section{DISCUSSION AND CONCLUSIONS}

As it was demonstrated in \S3, X-ray and $\gamma$-ray observations of blazars,
combined with our best guesses regarding the central environments
in quasars, allow us to estimate the leptonic  and total energy fluxes.
The former is found to be too small
to power the observed $\gamma$-ray flares or to support the energetics
of the radio lobes. Therefore, the energy flux in blazar jets
must be dominated by protons or magnetic fields,  but with the number of
e$^+$e$^-$-pairs greatly exceeding the  number of protons.
In two extreme cases we have: (A) a  proton dominated jet, with
a leptonic content 
$n_e/n_p \sim 20 (\eta_{diss}\eta_e/0.1)/(\bar \gamma_{inj}/10)$ (see Eq. 4); 
or (B) a  Poynting flux-dominated jet, with
$\sigma \simeq 10 L_{j,47}/[(\bar \gamma_{inj}/10)L_{e,45}]$
and  $n_e/n_p > m_p/m_e$.

The lack of soft X-ray excesses in blazar spectra, and of
soft X-ray precursors to non-thermal flares in blazar light curves ---
both  predicted to be produced by Comptonization of external radiation by cold
electrons in a jet (see \S4) ---
indicate that jet Lorentz factors are much smaller at distances
from the black hole $\la 10^3 R_g \sim 10^{17}$cm
than in the blazar zone.  
There is no convincing evidence that the jet acceleration continues
on larger scales: similar bulk Lorentz factors to those deduced from
blazar  models are inferred in parsec scale jets (Jorstad et al. 2001),
while bulk Lorentz factors deduced from modeling optical and
X-ray production in  kiloparsec-scale jets are at most comparable to
the blazar values (see, e.g., Scarpa \& Urry 2002; Tavecchio et al. 2004).
Hence, kinematical  data  from all spatial scales and the lack of
bulk  Compton radiation suggest that the jet acceleration saturates around
the blazar zone. In case  (A) above, this implies
that the conversion of Poynting flux to kinetic
energy  takes place
near or within the blazar zone. In case  (B),
the jet should stay strongly magnetically dominated over all scales.

The presence of hot spots ---  very compact features marking the
abrupt termination of relativistic jets --- as well as the
polarimetry of kiloparsec scale jets, seem to favor large scale
jets with $\sigma < 1$ and, therefore, case (A). This
leads us to speculate that the strongly enhanced activity of the
jet in the blazar zone may be related to the final stages of the
conversion of a magnetically dominated jet to a kinetic
energy-dominated jet. The dynamical events in blazars such as
flares can be driven by shocks, which are favored because of the
resulting broad energy distribution of electrons. However,
efficient dissipation and particle acceleration appear to require
strong shocks, and, therefore, $\sigma \ll 1$, and it is quite
unlikely that such $\sigma$ can be reached within the blazar zone
by an ideal MHD flow (Begelman \& Li 1994). This difficulty can be
overcome by postulating dissipative enhancement of the conversion
process. The magnetic dissipation can be stimulated and amplified
by kink and pinch instabilities, which  seem to be unavoidable
ingredients of Poynting flux-dominated  jets (Eichler 1993;
Begelman 1998). Alternatively, one can consider a direct
connection of the blazar activity with the magnetic dissipation
events, provided they can result in production of electrons with a
broad energy distribution. Since in this case the conversion to
$\sigma \ll 1$ is not required, the jets may propagate to larger
distances keeping  toroidal magnetic fields close to the
equipartition value.

In both shock and magnetic dissipation scenario of blazar
activity, inertia of the matter is dominated by protons and,
therefore, the jet is very likely to be powered by an accretion
disk rather than the magnetosphere of a Kerr black hole. It can be
loaded by protons evaporating from the disk surface or injected
into the disk corona by magnetic eruptions, and enriched  by pairs
created by high energy photons produced in localized  disk flares
(Stern et al. 1995), in the entire corona (Beloborodov 1999), or
following interaction of the proto-jet with the external radiation
field (Sikora \& Madejski 2000). Since for $n_e/n_p \gg 1$ the
effective radiation pressure can be super-Eddington even for
sub-Eddington accretion rates, the MHD  outflow can be launched
even for nearly vertical  magnetic field lines,  in contrast to
the $> 30$  degree tilt required for magnetocentrifugal
acceleration (Blandford \& Payne 1982). In this case, the outflow
can be driven mainly by radiation pressure within the sub-Alfvenic
region, while the main part of acceleration is provided by magnetic stresses
in the super-Alfvenic region and enhanced by dissipative processes supported
by MHD instabilities.

\smallskip
In summary, we conclude:

\noindent
$\bullet$ The kinetic
energy flux of leptons, estimated from the emissivity
of blazar  events, is too small to support energetics
of blazars and of radio lobes in quasars;

\noindent
$\bullet$ Studies of kinematics and dynamics of quasar jets indicate that
their power on the parsec- and kilo-parsec scales is likely
dominated by protons,
but the present data do not allow us to distinguish between the cases
$\sigma \la 1$ and $\sigma \ll 1$;

\noindent
$\bullet$
Dynamical events associated with
the blazar phenomenon and the lack of evidence
for acceleration of jets beyond the blazar zone
suggest that blazar activity can be related to
the final stages of the conversion of initially
Poynting flux dominated-jets into proton-dominated
jets. MHD instabilities may play a key role in this process;

\noindent
$\bullet$
Magnetic reconnection scenarios must be better understood
and quantitative theories for the acceleration of electrons in the presence of
protons --- both in shocks and in reconnection sites --- are needed to
determine the nature of dissipative events responsible for the blazar
activity;

\noindent
$\bullet$
Domination of the matter inertia by protons suggests
that accretion disks have the primary role in powering quasar  jets.
A large pair content --- deduced from the
emissivity and energetics of blazar events  and provided by
high energy processes in the hot accretion disk corona --- guarantees launching
of MHD outflow even in the case of nearly  vertical magnetic field lines.

\noindent
$\bullet$
As indicated by the lack of bulk-Compton
features in the spectra of blazars, acceleration of a jet takes at least
$10^3$ gravitational radii.

\acknowledgments

This project was partially supported by Polish KBN grants 5 P03D
00221, PBZ-KBN-054/P03/2001, by LEA Astro-PF grant  and
NSF grant AST-0307502,
by Chandra
grants no. GO1-2113X and GO4-5125X from NASA via Smithsonian
Astrophysical Observatory, and by the Department of Energy
contract to SLAC no. DE-AC3-76SF00515. M.S. thanks the Fellows of
JILA (Univ. of Colorado), SLAC (Stanford University), and IAP for
their hospitality. JPL was supported in part by a grant from
the CNRS GDR-PCHE.

\end{document}